\documentclass{article}
\usepackage{amsmath,graphicx,color}
\usepackage{algorithm}
\usepackage[noend]{algpseudocode}
\usepackage{subcaption}
\usepackage{spconf}
\usepackage{hyperref}

\usepackage{fancyhdr}

 \fancypagestyle{FirstPage}{
 \lfoot{\copyright 2020 IEEE. Personal use of this material is permitted. Permission from IEEE must be obtained for all other uses, in any current or future media, including reprinting/republishing this material for advertising or promotional purposes, creating new collective works, for resale or redistribution to servers or lists, or reuse of any copyrighted component of this work in other works.}
}

\title{IMPROVING THE CLASSIFICATION  OF RARE CHORDS WITH UNLABELED DATA}
\name{Marcelo Bortolozzo, Rodrigo Schramm, Claudio R. Jung\thanks{This work is partly supported by Coordenação de Aperfeiçoamento de Pessoal de Nível Superior (CAPES) –
Finance Code 001 and by the Brazilian Research Council (CNPq)}}
\address{Institute of Informatics at UFRGS, Porto Alegre, Brazil}
\begin{document}
%
\maketitle

\thispagestyle{FirstPage}

%
\begin{abstract}
In this work, we explore techniques to improve performance for rare classes in the task of Automatic Chord Recognition (ACR). We first explored the use of the focal loss in the context of ACR, which was originally proposed to improve the classification of hard samples. In parallel, we adapted a self-learning technique originally designed for image recognition to the musical domain.  Our experiments show that both approaches individually (and their combination) improve the recognition of rare chords, but using only self-learning with noise addition yields the best results.
\end{abstract}

\begin{keywords}
Automatic Chord Recognition, Data Imbalance, self-training, focal loss.
\end{keywords}
\section{Introduction}
\label{sec:intro}

The field of Music Information Retrieval (MIR) comprises a wide range of different tasks. In particular, Automatic Chord Recognition (ACR) aims to identify each musical chord  being played at a given instant on an audio segment.  The current state-of-the-art on ACR involves deep learning techniques~\cite{Pauwels:2019}, which require large amounts of labeled data. However, most publicly available chord datasets~\cite{harte2010towards, goto2002rwc, burgoyne2011expert}
consist of no more than a few hundreds of music recordings. Furthermore, they are inherently imbalanced, which generates a clear bias toward some of the more common chord types.

To show the degree of imbalance in a typical dataset, we present in Figure~\ref{fig:classimbalancefigure} the chord type distribution of the Beatles and Queen Isophonics Dataset \cite{harte2010towards}, which is one of the most popular datasets for ACR. We can clearly see that its composition is overwhelmed by major (maj)  and minor (min) chords, which together represent almost 80\% of all examples. Besides being the most common types of chords, they are also the easiest ones, as they are less ambiguous and are composed of fewer pitches.

Dealing with highly imbalanced data is a challenge when training a classifier for any given problem, as there will be very few (or sometimes no examples at all) for some of the classes, which hinders its accuracy or makes it impossible to learn at all. In the context of ACR, the problem is even more challenging since creating new chord datasets or expanding existing ones is not easy.  Unlike trivial tasks such as labeling common objects in images, the identification of chords requires musical training, and it is very time-consuming. A second issue with imbalanced data relates to the metrics chosen to evaluate the results, as a classifier might perform well on the common classes and very poorly on all others, but still get good scores if a global accuracy is used. 


This work focuses on improving the classification scores of uncommon chords. For that purpose, we explore a self-learning strategy that combines labeled and (large) unlabeled datasets. We show that our approach can significantly increase balanced accuracy metrics \cite{cho2014improved} with little (or no impact) to the global accuracy.

\section{Related Work}
\label{sec:related}

As reported in~\cite{Pauwels:2019}, the evolution of ACR techniques has moved from knowledge-driven to data-driven systems. Advance in the field has been promoted by new deep learning models,  
including hierarchical and multi-stage classification \cite{wu:li:2019} and alternative target label representation to provide better chord class separation \cite{carsault2019using, mcfee2017structured}.  However, chord vocabulary and associated balance of chord classes are still big issues for ACR. In \cite{jiang2019large} and~\cite{wu:li:2019}, a chord-specific framework is used in order to mitigate the difficulties in handling more complex chords. The approaches described in~\cite{mcfee2017structured, jiang2019large} use Encoder-Decoder architectures and chord structure decomposition in order to achieve better results for large vocabulary recognition. 

%
%
%
%
%
It is also important to point out that imbalanced data also arises in several other problems, such as image detection/classification and medical data processing. A systematic review of class imbalance in CNNs was presented in~\cite{buda2018systematic}, where authors claim that oversampling proved to be a good choice. However, ACR datasets present some classes with so few samples that direct oversampling might lead to overfitting. Another common strategy is to use weights to compensate for the class imbalance (so that less frequent classes present larger weights in the loss function). In fact, a combination of this strategy with a modification to the cross-entropy loss (called \textit{focal loss}) to decrease the weight of ``easy'' samples was presented in~\cite{lin2017focal}, with good results for object detection. In the context of image classification, recent approaches have shown that using  synthetic examples for rare classes can improve the classifier~\cite{beery2020synthetic}. 


 Finally, there is a recent trend of using self-learning approaches to boost the accuracy results in image classification problems. The noisy student approach presented in~\cite{xie2020self} combines labeled and unlabeled data with a class balancing scheme. Despite the good results shown in their paper, the extension for sequential problems (such as ACR) is not trivial, since the input is a sequence of chords (and not a single one). Recently, their approach was extended for speech recognition~\cite{park2020improved}, where some  considerations about sample selection on the unlabeled dataset for sequential data were made. 
 





\section{The Proposed Approach}
\label{sec:method}

This paper follows two main directions for improving the classification results of rare chords in ACR. First, we explore the use of class balancing at the loss function using a focal loss~\cite{lin2017focal}, accounting for sequential data. Second, we extend the noisy student approach in~\cite{xie2020self} using a large unlabeled dataset for ACR and a careful instance selection scheme tailored to sequential data.

\vspace{.2cm}
\noindent
\textbf{Focal Loss:}
as mentioned before, existing ACR datasets suffer from strong imbalance, and some of the most common chords are also easier to classify than others. The log-loss function,  used in most classifiers, assigns a relatively high value for these ``easy'' chords even for high classification probabilities, 
%
which lowers even more the relative importance of rare chords. To alleviate this problem, we tested the use of the  \textit{focal loss}~\cite{lin2017focal}, which introduces a power term to the traditional log-loss aiming to lower the loss value of easier samples:
\begin{equation}
FL = (1-p_t)^\gamma \log(p_t),
\end{equation}
where $p_t$ is the estimated probability for ground truth chord and $\gamma$ is a parameter that determines how much the weight of the loss of easy examples will be reduced.

\begin{figure}[tb]
\begin{subfigure}{.49\columnwidth}
    \centering
    \includegraphics[width=\linewidth]{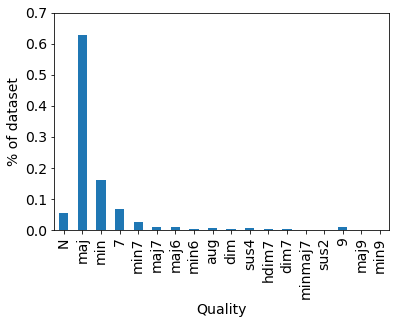}
    \caption{Isophonics}
    \label{fig:classimbalancefigure}
\end{subfigure}
\begin{subfigure}{.49\columnwidth}
    \centering
    \includegraphics[width=\linewidth]{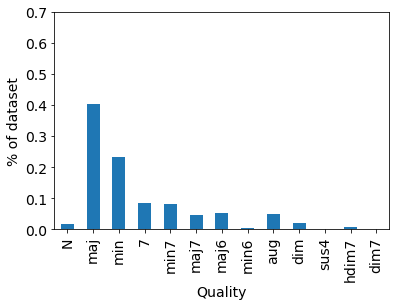}
    \caption{Balanced Pseudolabeled}
    \label{fig:pseudolabeldistribution}
\end{subfigure}
\caption{Comparison of chord type distributions on Datasets}
\end{figure}

\vspace{.2cm}
\noindent
\textbf{Self-Learning:}
the second strategy adopted in this paper was the use of large unlabeled datasets for improving the accuracy of the classifier, based on a state-of-the-art semi-supervised learning technique known as Noisy-Student~\cite{xie2020self}. The proposed method relies on using a classifier trained with a known dataset (teacher) to generate pseudo labels for a large unlabeled dataset. A subset of this unlabeled dataset is then selected, set aside and some augmentation techniques are applied to generate ``noisy'' samples. The original training dataset is then merged with this pseudo-labeled augmented dataset, and a new training dataset is obtained and used to train a new student model. This model then becomes the new teacher and is used to relabel the original unlabeled data, and the method proceeds iteratively in this way.

One important step in~\cite{xie2020self} is the procedure for selecting the subsets with pseudo-labels. It is done by randomly selecting the same amount of samples for each class, with replacement, from a dataset that contains labels predicted by the teacher that present confidence values larger than a threshold. Therefore the resulting subset is expected to be balanced and contain a sample of the most confident predictions.  This strategy has been tested on images and provided promising results, but when attempting to transfer it to the audio domain, in particularly ACR, additional care must be taken. First, while the original image-based application works in a single-input single-output manner, most ACR approaches classify a set of temporally adjacent chords (the input is a short audio sequence), which generates a set of labels that are temporally connected. 
Hence, it is not possible to pinpoint individual chords with high confidence, as they require the temporal context provided by temporally neighboring chords. Second, as we will choose chords along with their context, obtaining a truly balanced dataset is nearly impossible. Third, in the context of ACR, most of the examples are major and minor chords, and rarer chords represent a very small part of a dataset, as indicated in Figure \ref{fig:classimbalancefigure}. 




Having these issues in mind, we propose a new strategy for selecting the examples out of the pseudo-labeled dataset. First, we define \textbf{minLength} as the minimum length an audio excerpt needs to provide the required context for the classifier, and \textbf{desiredDuration}, which represents the total desired duration for each rare chord type across the dataset. This algorithm aims to obtain samples that together present around the same duration as the original labeled training dataset, so the \textbf{desiredDuration} is defined as the duration of that dataset divided by the number of rare chords present in the pseudolabels. For each rare chord class, we then select only excerpts with the corresponding label ordered by confidence, and iterate through them in that order. For each of the entries, we select an excerpt with duration \textbf{minLength} centered on the selected entry, or adjusting if at the beginning/end of the audio file. We then merge it with the new dataset, joining overlapping excerpts and adding its duration to \textbf{selectedDuration}. After the desired duration for this chord type is achieved, we move on to the next, until the dataset is complete.

The proposed strategy, summarized in Algorithm \ref{algorithm}, generates subsets that are much more balanced than a traditional ACR dataset, as can be seen in Figure \ref{fig:pseudolabeldistribution}. Major and minor chords still dominate, but that is inevitable since they tend to accompany the rare chords. It is also important to notice that the generated subset has a smaller variety of chords, as the classifier we used, a state-of-the-art  ACR technique based on the Transformer architecture \cite{park2019bi}, has a limited number of classes it is able to predict.



\begin{algorithm}
\small
\caption{Balanced Pseudolabel Dataset Selection}
\begin{algorithmic}[1]
\State \textbf{minLength} = minimum length for each audio excerpt\label{alg:step1}
\State \textbf{desiredDuration} = total duration desired for each rare chord type\label{alg:step2}
\For{each rare chord type \textbf{ct}}
    \State Filter labeled entries by \textbf{ct} ordering by prediction confidence
    \State Initialize \textbf{selectedDuration} = 0
    \For{each ordered \textbf{excerpt}}
        \State Create new excerpt centered on \textbf{excerpt}, with duration \textbf{minLength}
        \State Add it to dataset, merging overlaps to avoid duplicates
         \State Increment \textbf{selectedDuration} by duration of the excerpt
        \State If \textbf{selectedDuration} $\geq$ \textbf{desiredDuration} then move to next chord type
    \EndFor
\EndFor
\end{algorithmic}
\label{algorithm}
\end{algorithm}


Another important component of~\cite{xie2020self} is the addition of noise to the selected subset.  
For model noise, we maintained the recommended usage of dropout, and for input noise, we adapted it to augmentation techniques used in the context of MIR, which include pitch shifting and random audio excerpts from the UrbanSound8K dataset \cite{Salamon:UrbanSound:ACMMM:14}. We mixed the noise recorded from a city environment together with our pseudo-labeled audio examples using the MUDA library \cite{mcfee2015_augmentation}. This strategy adds a random component to the subset, and helps to reduce the presence of pitched artifacts that could corrupt the ground truth chord profiles. 

%

\section{Experimental Results}

We ran our experiments using the previously stated state-of-the-art classifier \cite{park2019bi} and implemented the necessary modifications in order to run both the focal loss and self-learning tests. For our labeled dataset, we used the Isophonics Queen and Beatles dataset \cite{harte2010towards}, and as our (large) unlabeled data, we used the audios indicated by the DALI dataset~\cite{meseguer2019dali} which results in around 5,000 songs without chord label annotations. We split the Isophonics dataset into training and validation sets using a 80/20 split, where the validation set was used for early stopping. For the testing stage, we used the RWC~\cite{goto2002rwc} chord dataset, yielding a cross-dataset experiment.

We generated the baseline results by running the unmodified version of the classifier through the pipeline. For the focal loss, we ran experiments with three different values of $\gamma$: 1, 2 and 5. For the Noisy-Student strategy, we ran two versions: one using only dropout and pitch shifting for augmentation, and another that also included random noise into the pseudolabeled dataset. Finally, we ran a final experiment that combined NoisyStudent with focal loss, configured with the value for $\gamma$ that provided the best preliminary results. The original paper for Noisy-Student obtained the best results by running three iterations, and  we  follow this recommendation in our work. 

\subsection{Evaluation Metrics}

A crucial issue observed in the context of ACR is that most songs are composed of only two chord types, so that many algorithms are able to obtain very good results by focusing on improving predictions for those types exclusively. This could happen when using one of the most common metrics in ACR, know as Weighted Chord Symbol Recall (WCSR)~\cite{harte2010towards}. It provides a global accuracy metric, but with scores proportional to the duration of each class, allowing for imbalanced results. The Chord Symbol Recall can be defined as:

$$
CSR = \frac{|S \cap S_t|}{|S_t|}
$$
where $S$ and $S_t$ are the predicted chord labels and the ground-truth labels, respectively. By weighing this value by track length we obtain the WCSR:

$$
WCSR = \frac{\sum{T_i * CSR_i}}{\sum{T_i}}
$$
where $T_i$ is the duration of the $i$th track.

As our goal in this work is to improve predictions for the more uncommon classes, it also required the usage of a different metric in order to properly measure our results. We decided to use a metric which would be capable of measuring how balanced the accuracy across the classes is. It is called Average Chord Quality Accuracy (ACQA) \cite{cho2014improved} and is defined as the sum of the WCSR weighed by chord type ($WCSR_C$) divided by the number of chord types present, and it penalizes predictions that are biased towards common chords. This WCSR weighed by chord is obtained as follows:

$$
WCSR_C = \frac{\sum{C_i * CSR_{Ci}}}{\sum{C_i}}
$$
where $C_i$ is the duration of the $i$th chord instance of type $C$ and $CSR_{Ci}$ its associated chord symbol recall. Finally we obtain the ACQA score as follows:

$$
ACQA = \frac{\sum{WCSR_C}}{\sum{|C|}}
$$
with $C$ being the set of chord types.



\begin{table}[]
\footnotesize
 \begin{center}
 \scalebox{0.9}{%
 \begin{tabular}{|c|c|c|c|}
  \hline
   & WCSR & ACQA   \\
  \hline
  Baseline & 44.7 & 17.4   \\
  \hline
  Focal loss ($\gamma = 2$)  & 44.7 & 19.2    \\
  \hline
  NoisyStudent without random noise & 44.2 & 23.0 \\
  \hline
  NoisyStudent with random noise & \textbf{44.8} & \textbf{23.4} \\
  \hline
  NoisyStudent + Focal loss ($\gamma = 2$) & 44 & 23.1 \\
  \hline
  
 \end{tabular}
 }
\end{center}
\vspace{-0.2cm}
 \caption{Weighted Chord Symbol Recall (WCSR) and Average Chord Quality Accuracy (ACQA) for each of the experiments performed. }
 \label{tab:results}
\end{table}

\subsection{Analysis of the Results}

In Table~\ref{tab:results} we can see the results obtained for each of the different experiments. As the NoisyStudent ran for three iterations, we are displaying the results for the one which obtained the best ACQA performance. For the focal loss we only obtained a gain in the ACQA compared to the baseline on the experiment with $\gamma = 2$, with the WCSR kept on the same level, while for other values of $\gamma$ there was a loss on both metrics. For the experiments with NoisyStudent, we obtained better results compared to the baseline for the ACQA on all three instances, and furthermore, surpassed it on both WCSR and ACQA when applying random noise. The combination of NoisyStudent with random noise and focal loss, had a good improvement on the ACQA metric, but was still inferior to the same experiment without focal loss.

\begin{figure}[tb]
\begin{subfigure}{.49\columnwidth}
    \centering
    \includegraphics[width=\linewidth]{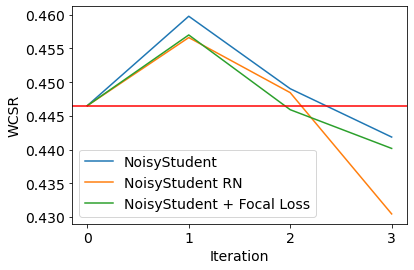}
    \caption{WCSR performance}
    \label{fig:wcsr}
\end{subfigure}
\begin{subfigure}{.49\columnwidth}
    \centering
    \includegraphics[width=\linewidth]{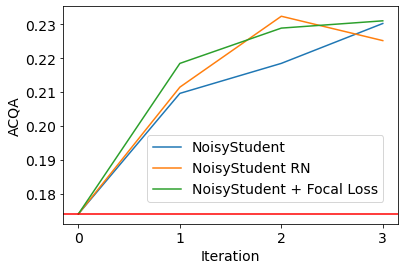}
    \caption{ACQA performance}
    \label{fig:acqa}
\end{subfigure}
\caption{Performance over the NoisyStudent, NoisyStudent with random noise (RN) and NoisyStudent with Focal Loss, compared to baseline (in red)}
\end{figure}

One relevant aspect in our results 
is the behavior obtained from the Noisy-Student variations used over each of its iterations. This can be seen in Figures \ref{fig:wcsr} and \ref{fig:acqa}, where we display the evolution of the WCSR and ACQA metrics starting from the baseline (Iteration 0) up to the third iteration.
The model initially obtain a better WCSR score on the first iteration by using Noisy-Student. However, it clearly starts to degrade after this point, becoming even worse than the baseline on the third iteration. This behavior is not reflected so clearly on the ACQA, where the results keep improving, or at least are maintained far above the baseline, which indicates that the increase of accuracy of rare chords on later iterations happened at the expense of the accuracy on the more common ones.

This trade-off between WCSR and ACQA had already been previously observed in \cite{deng2017large} in an even more radical way, where all gains obtained in the ACQA through the even chance sampling of the training data were accompanied by a reduction of the WCSR. They reported a gain of 20\% in the ACQA, versus a reduction of 3\% in the WCSR performance. On the other hand, our approach managed to -- up to some extent -- obtain an increase in both metrics, as was seen in the first two iterations of the Noisy-Student, with a gain of almost 3\% in the WCSR and one of 21\% for the ACQA, or managed to obtain an improvement on the ACQA by 33\% while maintaining the WCSR at the baseline level, as was seen in our experiment using Noisy-Student with random noise. 

Another way to observe this trade-off is by dividing the performance into chord classes. Table \ref{tab:chord-scores} shows the relative duration of some chord types in our training and testing datasets, along with the accuracy obtained on the different experiments. We can clearly see how the performance for the most common chords (maj and min) degrades with our attempts to improve the balance of our resulting predictions. However, one important aspect that should also be considered in this table is the great improvement achieved for very rare chord classes. For instance, the hdim7 chords, which composes less than 1\% of the Isophonics training dataset and obtained an accuracy of 2\% on the baseline, reaches an accuracy of more than 40\% when using the NoisyStudent self-learning technique, which starts with the same training dataset as the baseline. Although the other rare chord classes do not present such a large gain in their performance, they are all more accurate in our experiments than in the baseline. This happens in such a degree that the result obtained by the WCSR, as was presented, is kept at the same levels as the baseline, however with a much more balanced result, as now all rare chord types can be predicted more accurately.

\begin{table}[]
\footnotesize
 \begin{center}
 \scalebox{0.9}{%
 \begin{tabular}{|c|c|c|c|c|c|c|c|c|}
  \hline
   & maj & min & 7 & min7 & maj7 & dim & hdim7   \\
  \hline
  \hline
  \% of train dataset & 63 & 16.1 & 6.9 & 2.6 & 1 & 0.4 & 0.2    \\
  \hline
  \% of test dataset & 45 & 15 & 7.2 & 13.8 & 7.4 & 0.8 & 0.4    \\
  \hline
  \hline
  Baseline  & \textbf{61.6} & \textbf{69.3} & 31 & 14.9 & 1.5 & 11.8 & 2     \\
  \hline
  Focal loss & 58.5 & 63.3 & 30.7 & 27.1 &  7.5 & 20 & 8.5   \\
  \hline
  NoisyStudent & 53.5 & 52.4 & \textbf{48.5} & \textbf{45.5} & \textbf{11} & \textbf{28.9} & \textbf{41.3} \\
  \hline
   \end{tabular}
   }
\end{center}
\vspace{-0.2cm}
 \caption{Individual Chord accuracy associated to relative duration on Isophonics Dataset (training dataset) }
 \label{tab:chord-scores}
\end{table}


\section{Conclusion}
\label{sec:conclusion}

This paper presented two strategies for improving the recognition of rare chord types: the use of a focal loss to train a classifier, and the adaptation 
of a self-learning technique to deal with sequential data. Both strategies improved the overall in terms of the ACQA metric, but self-learning with noise presented the best results (even improving the WCSR scores).  However, we observed that in some cases, the more balanced accuracy comes as a result of a loss in the individual accuracy of common classes. Finally, our experiments also indicated that this technique has a very large potential for learning extremely rare examples, as we have demonstrated with chords that composed less than 1\% of the training data. 
Supplemental material can be found on the following repository: \url{https://gitlab.com/mcbortolozzo/chord-noisy-student}

\bibliographystyle{IEEEbib}
\bibliography{main}

\begin{thebibliography}{10}

\bibitem{Pauwels:2019}
J~Pauwels, K~O{\textquoteright}Hanlon, E~G{\'o}mez, and MB~Sandler,
\newblock ``20 years of automatic chord recognition from audio,''
\newblock in {\em International Society for Music Information Retrieval
  (ISMIR)}, Delft, Netherlands, 04/11/2019 2019.

\bibitem{harte2010towards}
Christopher Harte,
\newblock {\em Towards automatic extraction of harmony information from music
  signals},
\newblock Ph.D. thesis, 2010.

\bibitem{goto2002rwc}
Masataka Goto, Hiroki Hashiguchi, Takuichi Nishimura, and Ryuichi Oka,
\newblock ``Rwc music database: Popular, classical and jazz music databases.,''
\newblock in {\em Ismir}, 2002, vol.~2, pp. 287--288.

\bibitem{burgoyne2011expert}
John~Ashley Burgoyne, Jonathan Wild, and Ichiro Fujinaga,
\newblock ``An expert ground truth set for audio chord recognition and music
  analysis.,''
\newblock 2011.

\bibitem{cho2014improved}
Taemin Cho,
\newblock {\em Improved techniques for automatic chord recognition from music
  audio signals},
\newblock Ph.D. thesis, New York University, 2014.

\bibitem{wu:li:2019}
Y.~{Wu} and W.~{Li},
\newblock ``Automatic audio chord recognition with midi-trained deep feature
  and blstm-crf sequence decoding model,''
\newblock {\em IEEE/ACM Transactions on Audio, Speech, and Language
  Processing}, vol. 27, no. 2, pp. 355--366, 2019.

\bibitem{carsault2019using}
Tristan Carsault, Jérôme Nika, and Philippe Esling,
\newblock ``Using musical relationships between chord labels in automatic chord
  extraction tasks,'' 2019.

\bibitem{mcfee2017structured}
Brian McFee and Juan~Pablo Bello,
\newblock ``Structured training for large-vocabulary chord recognition.,''
\newblock in {\em ISMIR}, 2017, pp. 188--194.

\bibitem{jiang2019large}
Junyan Jiang, Ke~Chen, Wei Li, and Gus Xia,
\newblock ``Large-vocabulary chord transcription via chord structure
  decomposition.,''
\newblock in {\em ISMIR}, 2019, pp. 644--651.

\bibitem{buda2018systematic}
Mateusz Buda, Atsuto Maki, and Maciej~A Mazurowski,
\newblock ``A systematic study of the class imbalance problem in convolutional
  neural networks,''
\newblock {\em Neural Networks}, vol. 106, pp. 249--259, 2018.

\bibitem{lin2017focal}
Tsung-Yi Lin, Priya Goyal, Ross Girshick, Kaiming He, and Piotr Doll{\'a}r,
\newblock ``Focal loss for dense object detection,''
\newblock in {\em Proceedings of the IEEE international conference on computer
  vision}, 2017, pp. 2980--2988.

\bibitem{beery2020synthetic}
Sara Beery, Yang Liu, Dan Morris, Jim Piavis, Ashish Kapoor, Neel Joshi, Markus
  Meister, and Pietro Perona,
\newblock ``Synthetic examples improve generalization for rare classes,''
\newblock in {\em The IEEE Winter Conference on Applications of Computer
  Vision}, 2020, pp. 863--873.

\bibitem{xie2020self}
Qizhe Xie, Minh-Thang Luong, Eduard Hovy, and Quoc~V Le,
\newblock ``Self-training with noisy student improves imagenet
  classification,''
\newblock in {\em Proceedings of the IEEE/CVF Conference on Computer Vision and
  Pattern Recognition}, 2020, pp. 10687--10698.

\bibitem{park2020improved}
Daniel~S Park, Yu~Zhang, Ye~Jia, Wei Han, Chung-Cheng Chiu, Bo~Li, Yonghui Wu,
  and Quoc~V Le,
\newblock ``Improved noisy student training for automatic speech recognition,''
\newblock {\em arXiv preprint arXiv:2005.09629}, 2020.

\bibitem{park2019bi}
Jonggwon Park, Kyoyun Choi, Sungwook Jeon, Dokyun Kim, and Jonghun Park,
\newblock ``A bi-directional transformer for musical chord recognition,''
\newblock {\em arXiv preprint arXiv:1907.02698}, 2019.

\bibitem{Salamon:UrbanSound:ACMMM:14}
J.~Salamon, C.~Jacoby, and J.~P. Bello,
\newblock ``A dataset and taxonomy for urban sound research,''
\newblock in {\em 22nd {ACM} International Conference on Multimedia
  (ACM-MM'14)}, Orlando, FL, USA, Nov. 2014, pp. 1041--1044.

\bibitem{mcfee2015_augmentation}
B.~McFee, E.J. Humphrey, and J.P. Bello,
\newblock ``A software framework for musical data augmentation,''
\newblock in {\em 16th International Society for Music Information Retrieval
  Conference}, 2015, ISMIR.

\bibitem{meseguer2019dali}
Gabriel Meseguer-Brocal, Alice Cohen-Hadria, and Geoffroy Peeters,
\newblock ``Dali: A large dataset of synchronized audio, lyrics and notes,
  automatically created using teacher-student machine learning paradigm,''
\newblock {\em arXiv preprint arXiv:1906.10606}, 2019.

\bibitem{deng2017large}
Jun-qi Deng and Yu-Kwong Kwok,
\newblock ``Large vocabulary automatic chord estimation with an even chance
  training scheme.,''
\newblock in {\em ISMIR}, 2017, pp. 531--536.

\end{thebibliography}

\end{document}